# Robust Topological Terahertz Circuits using Semiconductors

Babak Bahari, Ricardo Tellez-Limon, and Boubacar Kante[*]

[*]bkante@ucsd.edu

*Department of Electrical and Computer Engineering, University of California San Diego, La Jolla, California 92093-0407, USA*

Topological Insulator-based devices can transport electrons/photons at the surfaces of materials without any back reflections, even in the presence of obstacles. Topological properties have recently been studied using non-reciprocal materials such as gyromagnetics or using bianisotropy. However, these effects usually saturate at optical frequencies and limit our ability to scale down devices. In order to implement topological devices that we introduce in this paper for the terahertz range, we show that semiconductors can be utilized via their cyclotron resonance in combination with small magnetic fields. We propose novel terahertz operating devices such as the topological tunable power splitter and the topological circulator. This work opens new perspectives in the design of terahertz integrated devices and circuits with high functionality.

The recent discovery of Topological Insulators intertwines the symmetry of crystals and the electronic topology to create intriguing interaction of wave and matter with topological order. Topological Insulators can transport electrons without any back-scattering at the surface of the materials even in the presence of obstacles. This phenomenon was shown for the first time in the Quantum Hall effect, where electrons in a magnetic field are transported without any back-scattering [1]. These systems have been extensively studied and properties such as immunity to disorder have been predicted and demonstrated. The recent demonstration by Raghu and Haldane [2] on the possibility to transfer some topological properties from fermionic to bosonic systems gives new degrees of freedom over the control of photons.

A number of robust waveguiding experiments have subsequently been proposed and demonstrated [3-10]. Applying an External Magnetic Field (EMF) to a 2D gyromagnetic Photonic Crystal (PhC), chiral boundary modes at microwave frequencies were induced as a result of time-reversal symmetry breaking in analogy with the integer Quantum Hall effect [6]. At higher frequencies where gyromagnetism is weak, the implementation of these ideas is challenging and 2D lattices with bianisotropy, temporal modulation, and ring resonators [4, 9], in both strongly and weakly coupled periodic arrays, have all been proposed.

Besides weak gyromagnetism, which results in devices with large footprint, material losses represent an issue to break the time-reversal symmetry at optical frequencies. Graphene was suggested to overcome this limitation and inhomogeneous strain can induce pseudo-magnetic fields [11].

In the terahertz spectral range, materials such as $Bi_2Se_3$, generating strong spin-orbit interaction, also protected by time-reversal symmetry, can be used [12, 13]. To the best of our knowledge, integrated devices have not yet been used for Topological Insulator applications in the terahertz spectral range. This frequency band $(0.1 - 30\,THz)$ is of interest for number of applications such as information and communication technologies, biology and medical sciences, non-destructive evaluation, and even global environmental monitoring [14].

A possible way to break the time reversal symmetry in this spectral range, is to use the cyclotron resonance effect [15-18] in semiconductor materials [19]. In these CMOS compatible materials, the cyclotron resonance can be triggered in the terahertz band with a small EMF [20].

In this paper, we report and numerically characterize topological circuits operating in the terahertz frequency range. Using the propagation of one-way modes at the boundaries of semiconductor photonic crystals due to the topological effect under a small magnetic fields, we propose a tunable topological power splitter and a topological circulator.

Materials containing free electrons can exhibit the cyclotron resonance effect, i.e, free electrons circle in the plane perpendicular to the direction of the applied magnetic field. The cyclotron frequency of the electrons is proportional to the strength of the EMF. This phenomenon results in a non-reciprocal interaction between the material and the electromagnetic wave propagating in forward and backward directions. Because of this effect, the relative permittivity tensor of the material presents off-diagonal terms with opposite signs [21].

The cyclotron resonance frequency, $\omega_c$, depends on the effective mass of the electrons, $m^*$, and the strength of the EMF, $B$, as $\omega_c = eB/m^*$, where $e$ is the charge of electron [21]. The dielectric permittivity tensor for a dispersive material exposed to an EMF in the $z$ direction can be described by [21]

$$\bar{\bar{\epsilon}}(\omega) = \epsilon_\infty \begin{bmatrix} \epsilon_{xx} & i\epsilon_{xy} & 0 \\ -i\epsilon_{xy} & \epsilon_{xx} & 0 \\ 0 & 0 & \epsilon_{zz} \end{bmatrix}, \tag{1}$$

with

$$\epsilon_{xx} = \left(1 - \frac{\omega_p^2}{\left(\omega + \frac{i}{\tau}\right)^2 - \omega_c^2}\right) \times \left(1 + i\frac{1}{\tau\omega}\right),$$

$$\epsilon_{xy} = -\frac{\omega_p^2}{\left(\omega + \frac{i}{\tau}\right)^2 - \omega_c^2} \times \frac{\omega_c}{\omega}, \tag{2}$$

$$\epsilon_{zz} = \left(1 - \frac{\omega_p^2}{\left(\omega + \frac{i}{\tau}\right)^2 - \omega_c^2}\right) \times \frac{\left(\omega + \frac{i}{\tau}\right)^2 - \omega_c^2}{\omega\left(\omega + \frac{i}{\tau}\right)},$$

where $\epsilon_\infty$ is the high frequency permittivity, $\omega_p$ the bulk plasma frequency, and $\tau$ the decay time characterizing material loss.

In metals, in order to have a cyclotron frequency, $\omega_c$, comparable with the plasma frequency, $\omega_p$, it is necessary to apply very large EMF (in the order of several Teslas). For small EMF, $\omega_c$ is negligible (Eqs. (1) and (2)), hence $\varepsilon_{xy} \approx 0$ and non-reciprocity is inexistent. This represents a limitation for on-chip devices.

However, due to the smaller effective mass of electrons in semiconductors, it is possible to obtain high $\omega_c$ comparable to $\omega_p$ at THz frequencies by applying a reasonable EMF [19]. For instance, Indium Antimonide ($InSb$) is a semiconductor with small energy gap of $0.17\ eV$ and a large electronic mobility ($\sim 7.7 \times 104 cm^2 V^{-1} s^{-1}$) [22], that has been used in the THz frequency range [19].

In this work, we used the dispersive relative permittivity of $InSb$ with very small loss and material parameters at room temperature given by $m^* = 0.014 m_0$ ($m_0$ is the free electron mass in vacuum), $\omega_p = 1.26 \times 10^{12}\ Hz$, and $\epsilon_\infty = 15.68$ [23, 24].

We consider the periodic array of $InSb$ rods with square lattice shape depicted in Fig. 1(a). The rods are invariant along the $z$ direction, and have a periodicity and a radius of $a_1 = 140\ \mu m$ and $r_1 = 0.25 a_1$, respectively.

Breaking the time-reversal symmetry can result in one-way edge modes that depend on the topological properties of the Bloch bands. These topological properties are characterized by the Chern number, that for the $n$-th band, is given by [2]

$$C_n = \frac{1}{2\pi i} \int \left(\frac{\partial \mathcal{A}_y^{nn}}{\partial k_x} - \frac{\partial \mathcal{A}_x^{nn}}{\partial k_y}\right) d^2 k, \tag{3}$$

where $\mathcal{A}$ is called *Berry connection* defined as

$$\mathcal{A}^{nn'}(\boldsymbol{k}) = \langle E_{nk} | \nabla_k | E_{n'k} \rangle, \tag{4}$$

where the brackets denote spatial integration over a unit cell.

In the calculated band diagram of Fig. 1(b), we do not observe a complete band gap. Reference [2] demonstrated that a band gap with non-zero Chern number can be opened by breaking the time-reversal symmetry of a gapless band diagram formed by two modes linearly touching each other (Dirac-shaped modes). Reference [3] showed that it is also possible to open a band gap with non-zero Chern number using quadratically crossing modes.

Here we demonstrate that none of these conditions are necessary. In general, it is only required to have a closed gap in entire Brillouin zone, regardless of the shape of the modes in the band diagram. For example, in Fig. 1(b), the second and third modes (red curves) close the band gap without touching each other. In the following, we will show that it is possible to open a band gap with non-zero Chern number using these two modes.

The Chern number is zero for all the modes in Fig. 1(b) since time-reversal symmetry has not been broken [25].

By applying an EMF of $B = 0.5\ T$ to the structure in the $z$ direction, the free electrons of $InSb$ rotate in the $xy$ plane with a frequency $\omega_c$, breaking the time reversal symmetry. According to Eqs. (1) and (2), the relative permittivity tensor becomes asymmetric. The resulting band diagram presents a band gap between the second and third modes (red curves in Fig. 1(c)).

Once the band gap is opened, the Chern number below the gap is a non-zero integer number ($\Delta C = +1$). Since the sum of the Chern numbers on the entire band diagram is invariantly zero [25], the Chern number above the gap is $-\Delta C = -1$. The resulting periodic structure is thus a non-trivial mirror that allows the propagation of edge modes only in one direction. The integration of this structure with a trivial mirror such as PEC, will lead to guided electromagnetic waves propagating in one direction and immune to any back reflection. To corroborate this one-way propagation, we placed a PEC close to the designed PhC to form a waveguide (inset in Fig. 2(a)). The waveguide is excited with a point source ($S$) located at the middle.

Fig. 2(a) shows the spectrum of the normalized power transmitted from the point source to the ports. The shaded area indicates the band gap region of the PhC shown in Fig. 1(c). We can observe an isolation between the ports in this region. This isolation is corroborated in the intensity maps at the frequencies $f_1 = 1.11\ THz$ (Fig. 2(b)) and $f_2 = 1.125\ THz$ (Fig. 2(c)), within and out of the band gap, respectively.

Due to its topological dependence, this edge mode features unidirectional robust propagation, ensuring the absence of back reflections due to defects or disorders. To illustrate this robustness, we studied the one-way propagation in a structure with sharp bends (Fig. 3(a)). For this structure, we used a PhC mirror instead of a PEC. The PhC mirror consists of a square lattice shape of $InSb$ rods. The periodicity and radius of the rods are $a_2 = 75.5\ \mu m$ and $r_2 = 0.35a_2$, respectively, with a plasma frequency of $\omega_p = 1.26 \times 10^{11}\ Hz$. With these parameters, the sum of the Chern number of the modes below the bandgap is zero (see Supplementary Information). This PhC can thus be used as a trivial mirror. This periodic structure was designed such that its band gap overlaps with the band gap of the PhC non-trivial mirror previously studied. The spectra of the normalized power transmitted from the point source to each port, are depicted in Fig. 3(b), showing the isolation between the ports. The robustness of the edge mode is clearly observed in the intensity map in Fig. 3(c) at a frequency within the band gap ($f_1 = 1.11\ THz$).

Based on these results, we propose a unidirectional power splitter formed by the combination of trivial (purple rods) and non-trivial (greed rods) PhC mirrors, as depicted in Fig. 4(a). The point source is placed between two mirrors to excite the one-way edge mode propagating in the indicated direction (black arrow). The power splits into two edge-modes at the crossing point between the four edge waveguides. The first one propagates clockwise toward port 1 (around the top-right PhC), while the second mode propagates counterclockwise toward port 2 (around the bottom-right PhC). Therefore, due to this asymmetric propagation, the amount of power flowing to each port is different ($\kappa_1$ and $\kappa_2$).

Both, $\kappa_1$ and $\kappa_2$, can be controlled by varying the radius of a defect ($InSb$ rod) placed at the center of the power splitter (see Supplementary Information). This first approach is just a static control, since for a fixed radius of the defect the power flowing to the ports will remain unchanged.

We can provide a dynamic control of the power splitter by changing the phase of a control signal ($C$) injected from the fourth remaining channel (left port), as illustrated in Fig. 4(a). This control signal constructively or destructively interferes with the fixed source signal ($S$), modifying the propagation of the edge mode toward the ports. In Fig. 4(b), we show the normalized power at each port as a function of the phase of the control signal, $\Phi$. For example, for $\Phi = 0\ rad$, the power is equally transmitted to each port (Fig. 4(c)). For $\Phi = 1.57\ rad$ the mode propagates toward port 1 (Fig. 4(d)), while for $\Phi = 4.71\ rad$, the mode is mainly propagating toward port 2 (Fig. 4(e)).

Following the previous results, we present a topological circulator operating at terahertz frequencies (Fig. 5(a)). The device consists of a symmetric array of trivial (purple rods) and non-trivial (green rods) PhC mirrors of $InSb$, with four ports. At the corners of the central PhC mirror –core of the circulator– four defects of radius $r = 80\ \mu m$ (orange rods) are placed to control the power flowing toward each port. The circulator is excited from port 1 with a point source at a frequency $f = 1.11\ THz$. As can be observed from the intensity map of Fig. 5(b), the unidirectional edge mode is mainly transmitted to port 2. The isolation between port 2 and 3 is 15.8 $dB$, and between port 2 and 4 is 18.2 $dB$.

We must remark that the number of channels can be modified by changing the geometrical shape of the core, such as triangular or pentagonal to operate with three or five ports.

In conclusion, we demonstrated topological terahertz devices using the one-way propagation of edge modes in periodic structures based on the concept of topological electromagnetic insulator. By applying a small EMF and breaking the time-reversal symmetry of $InSb$ semiconductor, through its cyclotron resonance in the terahertz frequency range, a variety of topological devices can be realized. We also demonstrated that in order to open a non-trivial gap between two modes, it is not necessary to have the modes touch each other either linearly or quadratically in the band diagram.

We proposed a power splitter able to dynamically tune the amount of power flowing to the ports, by varying the phase of a control signal. Based on the robustness of the power splitter, we also proposed a topological circulator operating at terahertz frequencies.

Due to their material properties and compatibility with CMOS manufacturing technology, semiconductors open new perspectives in the design of a new generation of integrated circuits with high functionality based on topological effects.

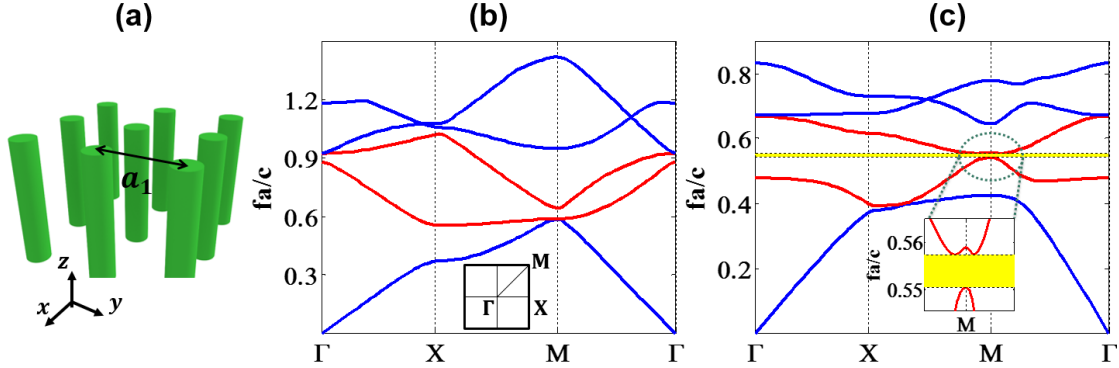

FIG. 1 (a) Schematic representation of a square lattice periodic array of $InSb$ rods. The periodicity and radius of the rods are $a_1 = 140\ \mu m$ and $r_1 = 0.25a_1$, respectively. (b) Band diagram for the periodic array along the irreducible Brillouin zone depicted in the inset. (c) By applying an EMF of $B = 0.5\ T$ to the structure along the $z$ direction, a band gap is opened in the frequency range $f = [1.106, 1.118]\ THz$.

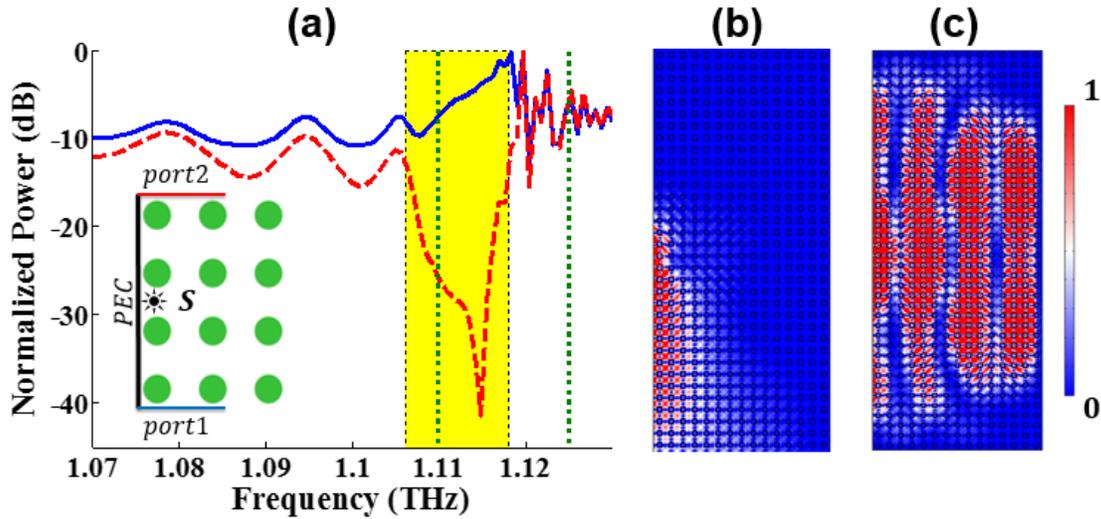

FIG. 2 (a) Normalized power at ports 1 (blue) and 2 (red) from a point source ($S$) placed in the middle of the edge waveguide between the PEC and PhC mirrors (inset). Intensity maps at (b) $f_1 = 1.11\ THz$ (within the band gap) and (c) at $f_2 = 1.125\ THz$ (out of the band gap). The edge mode is unidirectional for frequencies within the band gap region.

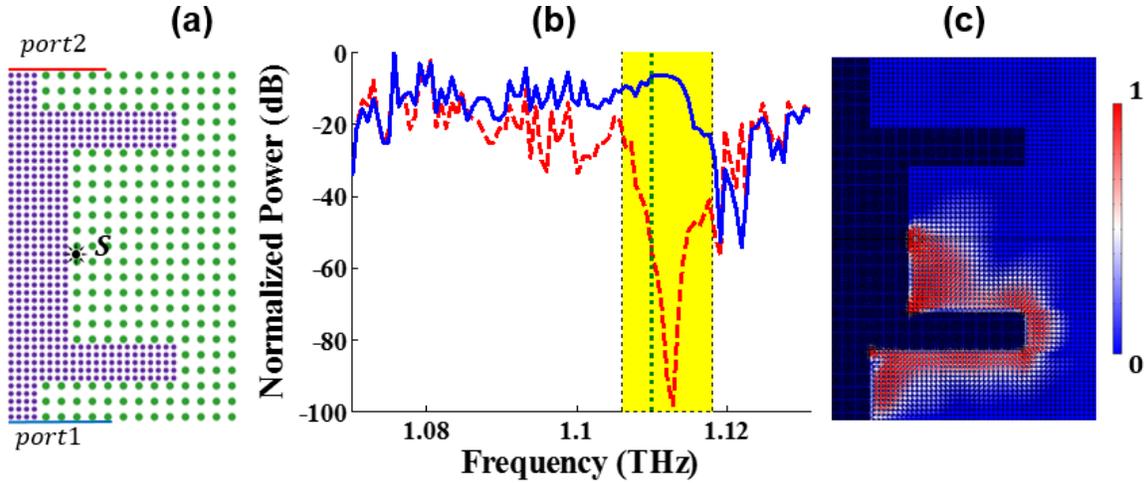

FIG. 3 (a) Schematic representation of a bended edge waveguide formed by trivial (purple) and non-trivial (green) PhC mirrors of $InSb$. The periodicity and radius of the trivial mirror are $a2 = 75.5\ \mu m$ and $r_2 = 0.35a_2$, respectively, and for the non-trivial mirror are the same than Fig. 1. (b) Normalized power at port 1 (blue) and 2 (red). (c) The intensity map at $f = 1.11\ THz$ (within the band gap region), shows the robustness of the one-way edge mode.

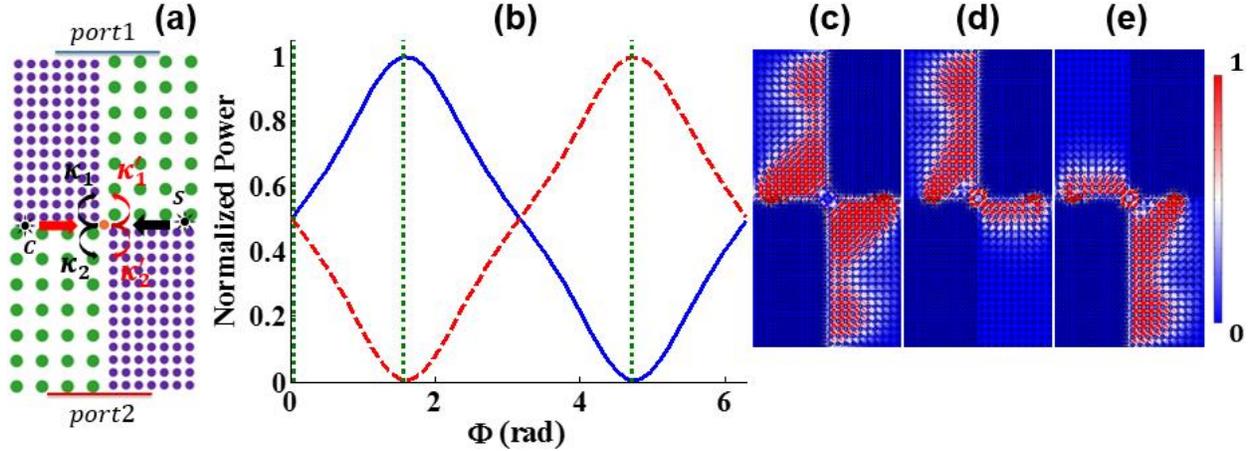

FIG. 4 (a) Schematic representation of a dynamic power splitter formed by two trivial (purple) and two non-trivial (green) PhC mirrors. A defect rod (orange) of radius $r = 84.8\ \mu m$ is placed at the center of the four edge waveguides. Two punctual sources are used as source ($S$) and control ($C$) signals (b) Normalized power at ports 1 (blue) and 2 (red) as a function of the phase variation of the control signal at $f = 1.11\ THz$. (c, d, e) Intensity maps for phases $\Phi = \{0, 1.57, 4.71\}\ rad$, respectively.

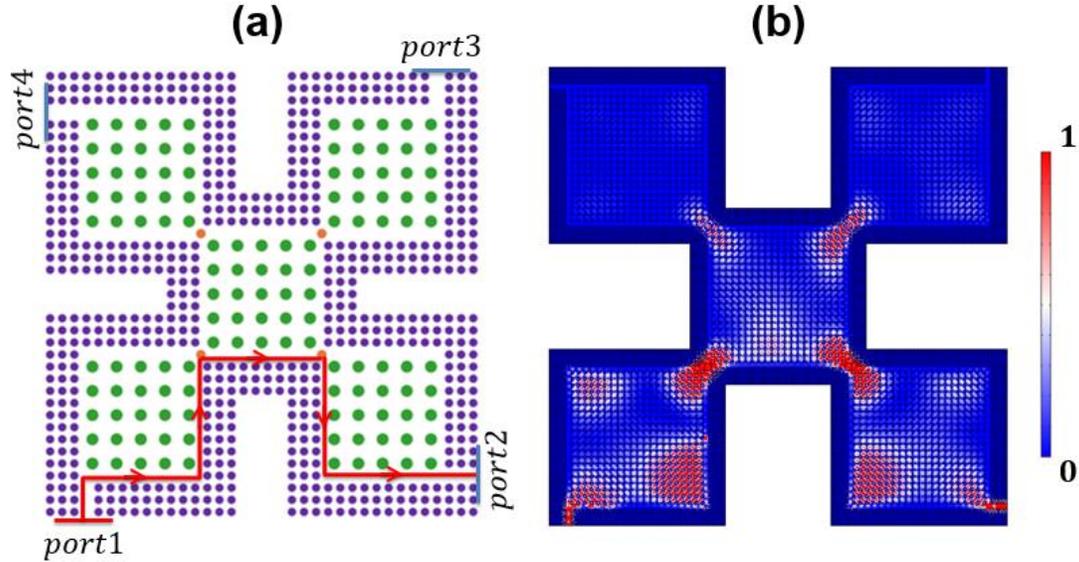

FIG. 5. (a) Schematic representation of a 4-port THz circulator formed by trivial (purple) and non-trivial (green) PhC mirrors, including four defect-rods (orange) of radius $r = 80\ \mu m$. The red arrows illustrate the unidirectional path of the edge mode. (b) The intensity map at $f = 1.11\ THz$, shows the power mainly flowing toward port 2 rather than 3 or 4.

# Supplementary Material

# Robust Topological Terahertz Circuits using Semiconductors

B. Bahari, R. Tellez-Limon, and B. Kanté

*Department of Electrical and Computer Engineering, University of California San Diego, La Jolla, California 92093-0407, USA*

**Design of the trivial mirror**

In order to design realistic topological devices, we designed a Photonic Crystal (PhC) mirror from the same semiconductor material ($InSb$) than the topological non-trivial mirror. The characteristic of this trivial mirror is that it can also operate in the presence of the External Magnetic Field (EMF), which is necessary to break the time-reversal symmetry in the non-trivial mirror.

The numerical results through all this paper, were obtained using the Finite Element Method (FEM) software COMSOL Multiphysics.

The PhC consists of a square lattice of $InSb$ rods (inset in Fig. S1(a)) with periodicity, radius of rods, and plasma frequency of $a_2$, $r_2 = 0.35a_2$, and $\omega_p = 1.26 \times 10^{11}\ Hz$, respectively. Fig. S1(a) shows the band diagram of the periodic structure for a periodicity of $a_2 = 75.5\ \mu m$. As we can observe, there is a small band gap between the first and second modes. Since there is no EMF, the Chern number is zero for the modes.

By applying an EMF to the structure, the time-reversal symmetry is removed, and the calculated band diagram is shown in Fig. S1(b). As depicted in the shaded region, the band gap between the first and second modes is enlarged, covering the frequencies range from $f_1 = 1.07\ THz$ to $f_2 = 1.17\ THz$. This frequency range covers the operating frequency range of the non-trivial mirror, which is between $f'_1 = 1.106\ THz$ and $f'_2 = 1.118\ THz$.

The Chern number of the modes below the band gap in Fig. S1(b), is zero as well. The only difference with the diagram of Fig. S1(a), is that the band gap is enlarged when the EMF is applied. Therefore, the opened band gap is not due to the difference in the Chern number. As a result, the PhC operating in the frequency range of the band gap is a trivial mirror.

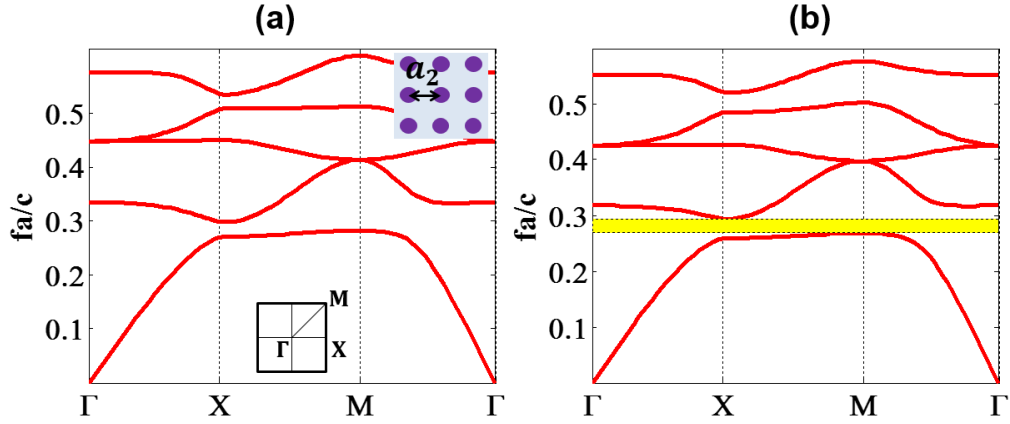

FIG. S1 (a) Band diagram for the periodic array of $InSb$ rods with periodicity and radius $a_2 = 75.5\ \mu m$ and $r_2 = 0.35a_2$, respectively (top inset), along the irreducible Brillouin zone (bottom inset). (b) The band gap is enlarged in the frequency range $f = [1.07, 1.17]\ THz$ by applying an EMF of $B = 0.5\ T$ to the structure.

## Power splitter

As explained in the manuscript, we can design a unidirectional power splitter by using a combination of trivial and non-trivial mirrors. The device, shown in Fig. S2(a), consist of two trivial PhC mirrors (purple rods) and two non-trivial PhC mirrors (green rods) symmetrically opposed by the center, forming four edge waveguide channels. The end facet of the top and bottom channels are labeled as port 1 and 2, respectively. A point source ($S$) is placed between the two mirrors in the right channel to excite the one-way edge mode propagating in the direction indicated by the black arrow.

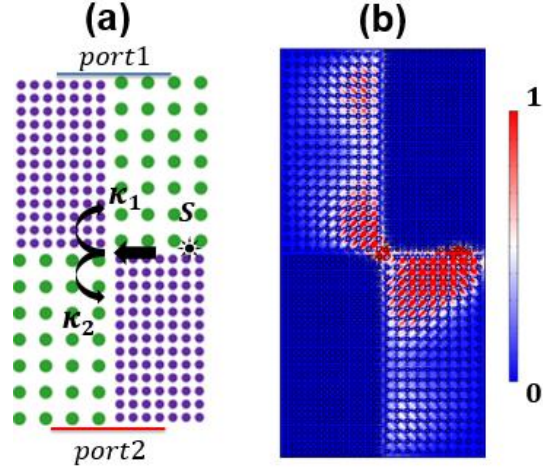

FIG. S2 (a) Schematic representation of a power splitter formed by two PhC trivial mirrors (purple) and two non-trivial mirrors (green) symmetrically opposed by the center. The one-way edge mode excited from the right channel with a point source ($S$), is split at the center of the structure, propagating with different amounts of power toward port 1 and 2. (b) Intensity map of the power splitter for a frequency $f = 1.11\ THz$.

The power is split into two edge modes at the crossing point between the four mirrors. The first one propagates clockwise toward port 1, while the second mode propagates counterclockwise toward port 2. Because of this asymmetric propagation, the amounts of power flowing to port 1($\kappa_1$) and to port 2 ($\kappa_2$) are different. This situation can be observed in the intensity map of Fig. S2(b) computed at a frequency $f = 1.11\ THz$ (within the band gap).

The difference between the power arriving at the two ports can be controlled by varying the radius of a defect ($InSb$ rod) placed at the crossing point between the four mirrors, as represented with the orange rod in Fig. S3(a). This dependence is observed in Fig. S3(b), where we plotted the normalized power at port 1 (blue) and port 2 (red) as a function of the radius of the defect. These results were computed at $f = 1.11\ THz$ (within the band gap region).

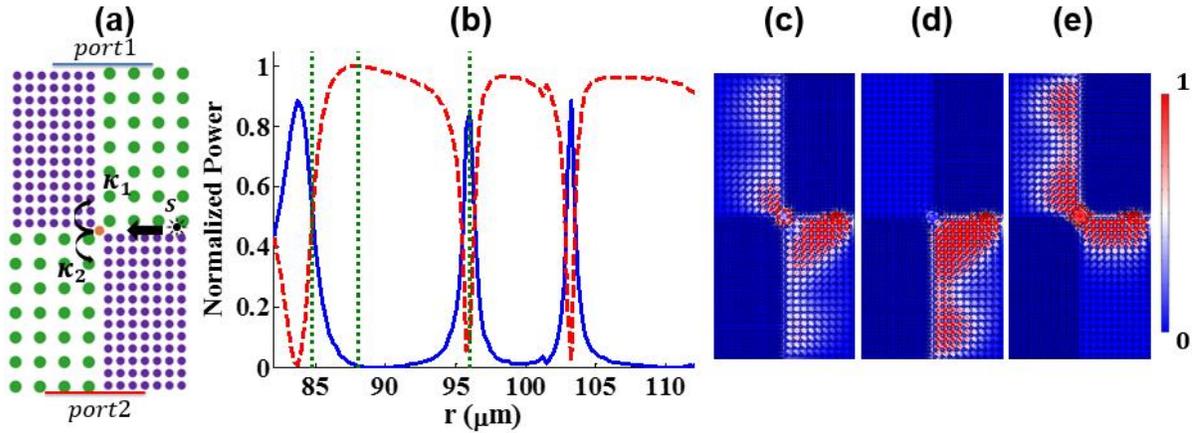

In Fig. S3(c-e) we show the intensity of the edge mode for three different radii of the defect, at the positions of the three green dotted vertical lines in Fig. S3(b). For a radius of 84.8 $\mu m$, the power is equally transmitted to both ports (Fig. S3(c)), while for radii of 88 $\mu m$ and 96 $\mu m$, the power is transmitted toward port 2 (Fig. S3(d)) and port 1 (Fig. S3(e)), respectively.

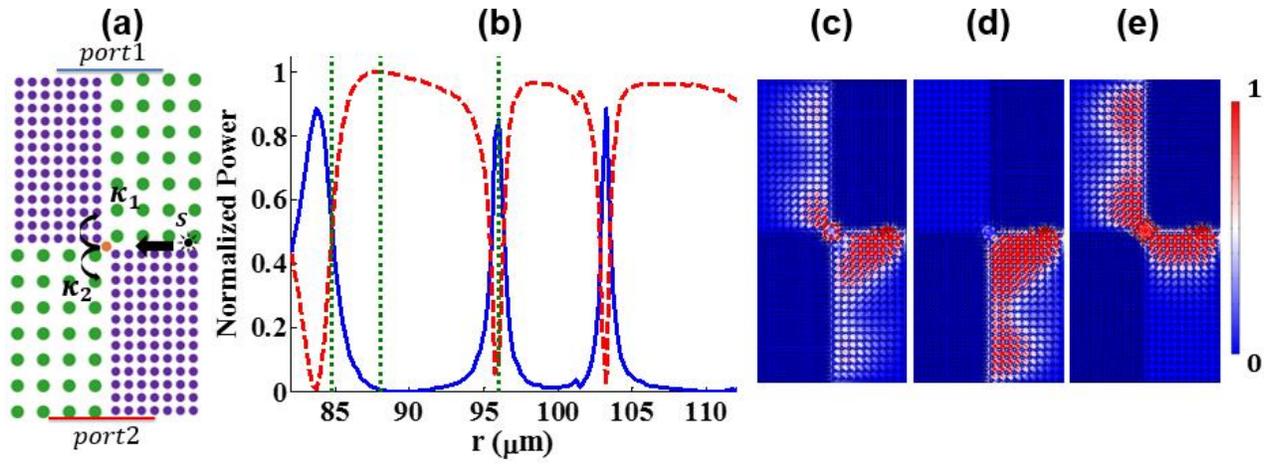

FIG. S3 (a) Schematic representation of a power splitter formed by trivial (purple) and non-trivial (green) PhC mirrors of $InSb$ containing a defect (orange rod) at the center. (b) Normalized power at ports 1 (blue) and 2 (red) as a function of the radius, $r$, of the defect for a frequency $f = 1.11\ THz$. (c, d, e) Intensity maps for $r = \{84.8, 88, 96\}\ \mu m$, respectively.